\documentclass[prl,aps,showpacs,twocolumn,superscriptaddress]{revtex4-1}
\usepackage{graphicx}
\usepackage[T1]{fontenc}
\usepackage{amssymb,latexsym,amsthm,amsmath} 
\usepackage{color}
\usepackage{listings}
\usepackage{enumerate}
\usepackage{changes}
\begin{document}

\title{Assortativity and leadership emerge from anti-preferential attachment in heterogeneous networks}

 \author{I. Sendi\~na-Nadal}
 \affiliation{Complex Systems Group, Universidad Rey Juan Carlos, 28933 M\'ostoles, Madrid, Spain}
 \affiliation{Center for Biomedical Technology, Universidad Polit\'ecnica de Madrid, 28223 Pozuelo de Alarc\'on, Madrid, Spain}
 \author{M. M. Danziger}
 \affiliation{Department of Physics, Bar Ilan University, Ramat Gan 52900, Israel}
 \author{Z. Wang}
 \affiliation{School of Automation, Northwestern Polytechnical University, Xi'an 710072, China}
\affiliation{Interdisciplinary Graduate School of Engineering Sciences, Kyushu University, Fukuoka, 816-8580, Japan}
 \author{S. Havlin}
 \affiliation{Department of Physics, Bar Ilan University, Ramat Gan 52900, Israel}
 \author{S. Boccaletti}
 \affiliation{CNR- Institute of Complex Systems, Via Madonna del Piano, 10, 50019 Sesto Fiorentino, Florence, Italy}
 \affiliation{The Italian Embassy in Israel, 25 Hamered st., 68125 Tel Aviv, Israel}

\begin{abstract}
Real-world networks have distinct topologies, with marked deviations from purely random networks.  
Many real-world networks exhibit degree-assortativity, with nodes of similar degree more likely to 
link to one another. Though microscopic mechanisms have been suggested for the emergence of other 
topological features, assortativity has prove{n} elusive.  Though assortativity can be artificially 
implanted in a network via degree-preserving link permutations, this destroys the graph's hierarchical 
clustering and does not correspond to any microscopic mechanism. Here, we propose the first generative 
model which creates heterogeneous networks with scale-free-like properties {in degree and clustering distributions} and tunable realistic assortativity.  
Two distinct populations of nodes are {incrementally} added to an initial {network by selecting a subgraph to connect to at random.} {One population} (the followers) {follows} 
preferential {attachment}, {while the other population} (the potential leaders) connect{s} via {\textit{anti-preferential} attachment: they link to} lower degree nodes {when added to the network}.  By selecting the lower degree nodes, the potential leader nodes maintain high {visibility} during the growth process, eventually growing into hubs. The evolution of links 
in Facebook empirically validates the connection between the initial anti-preferential attachment and long 
term high degree. {In this way}, our work sheds new light on the structure and evolution of social networks.
\end{abstract}


\maketitle

\section*{Introduction}
Networks with scale-free(SF)-like degree distributions represent a
wide range of systems\cite{Albert2000,Barabasi2002, Wuchty2003,Newman2003b, Boccaletti2006, Boccaletti2014}. 
The topology of real-world networks (RWNs) {often features} deviations from a pure power-law distribution{\cite{Barabasi2002}} { $P_k\sim k^{-\gamma}$}, together with {hierarchical clustering}{\cite{Ravasz2002} $C_k\sim k^{-\omega}$}. {One ubiquitous feature of many} RWNs {is} degree-degree
correlations: {two nodes are more likely to be linked } to one another if they are of similar {(assortative)} or dissimilar {(disassortative)} degree. 
Assortativity is generally found in
social and collaboration RWNs, while disassortativity is common in technological and biological RWNs\cite{Newman2002,Newman2003}.

SF networks have been studied in the context of generative models, and
simple rules relating to the formation of new links have been shown to
lead to power-law degree distributions with
non-hierarchical\cite{Boccaletti2007,Brede2011} and
hierarchical\cite{Barabasi1999,Dorogovtsev2000,Vazquez2003,Chung2003,Krapivsky2005,Clauset2008,Lorimer2015}
traits. Static SF network models\cite{Bender1978} have also been
proposed with controlled assortativity\cite{Zhou2008, Bassler2015}, and
growing SF networks have been studied  with assortative\cite{Catanzaro2004,Quayle2006,Toivonen2006,Tran2010,Shang2014},
disassortative\cite{Boccaletti2007,Subelj2013} and both types\cite{Brede2011} of degree mixing.

\begin{figure}
  \centering
 \includegraphics[width=\linewidth]{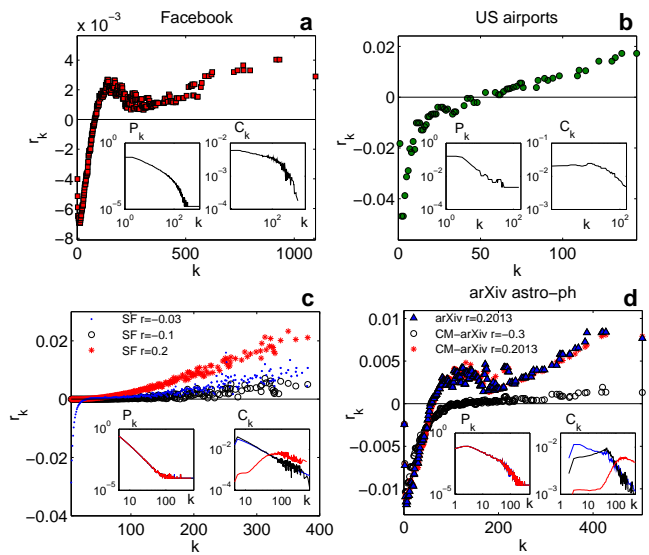}
\caption{ \textbf{ Local assortativity $r_k$ {\it vs.} the node degree $k$ for real\cite{konect} and artificial networks.} (\textbf{a})
 Data from friendships of Facebook users\cite{facebook-data} ($N=63,392$, $L=816,886$,
  $\langle k \rangle =26$, $r=0.1768$). (\textbf{b})  Network of the 500 busiest commercial airports in the United States\cite{Colizza2007}.
A tie exists between two airports if a flight was scheduled in
2002. ($N=500$, $L=2,980$, $\langle k \rangle =11.92$, $r=-0.2678$).
(\textbf{c}) Random SF networks ($N=10,000$, $\langle k\rangle =10$) with almost neutral ($r=-0.03$, blue dots), disassortative ($r=-0.1$, black circles) and assortative ($r=0.2$, red stars) mixing. (\textbf{d}) The Authors' collaboration graph from the
arXiv's Astrophysics section\cite{konect:leskovec107} ($N=17,903$, $L=196,972$, $\langle k \rangle =22 $, $r=0.2013$). Together with the real data (blue triangles),  $r_k$ is reported for
a CM reproducing the real degree sequence, after classical permutation methods have been applied, imposing the same $r$ value observed in the real network (red stars) and a negative ($r=-0.3$) value  (black circles).
Insets in panels (\textbf{a}-\textbf{d}) show the log-log plots of the
degree distributions $P_k$ and clustering coefficient $C_k$. \label{fig:realdata}
}
\end{figure}

In particular, a wide range of RWNs {feature } assortativity\cite{konect}, including online social\cite{facebook-data},
and neural\cite{Teller2014,deSantos2014} networks.
As it reflects a basic {\it birds of a feather flock together} property, it is not surprising that it is so ubiquitous. Rather, what is really surprising is that the contributions of different nodes to the graph assortativity level $r$ strongly depend on the degree.
Decomposing the assortativity spectrum, one can indeed describe the
\textit{local} {assortativity or} assortativeness\cite{piraveenan2008} $r_k$  of each
set of nodes with a given degree $k$ (see the Methods section). Many RWNs have a  pronounced {local}
 maximum in $r_k$ located near (but above) the average degree $\langle k \rangle$. In social networks such a feature even appears to be generic, while in technological and biological networks the maximum is less pronounced or even entirely absent.
 In Fig.~\ref{fig:realdata} {we show }
the qualitative difference in the inherent patterns of $r_k$ between
typical social networks (the friendship structure of Facebook
users\cite{facebook-data} and the Authors' collaboration graph from the arXiv's Astrophysics section\cite{konect:leskovec107,Whitfield2008,Eom2014}) and a technological one (the flights connecting the 500 busiest commercial airports in the United States\cite{Colizza2007}).

\section*{Results}
\subsection*{Empirical observations.}
The way traditional methods imprint assortativity into pre-generated networks is via degree-preserving link permutations\cite{Newman2003,Xulvi-Brunet2004}. {This approach yet presents a number of problems. On the one hand,} generating a graph with an ad-hoc
imprinted SF distribution {(Fig.~\ref{fig:realdata}c)} and {then} rewiring connections {\it does not}
yield the observed pattern of local assortativity, {on} the other {hand}, even starting from a configuration model (CM) retaining the original degree distribution\cite{Bender1978}, this procedure is only able to reproduce the real
assortativity pattern at the expense of destroying the other
significant features, {such} as the hierarchical inherent structure of
clustering (Fig.~\ref{fig:realdata}d and its bottom-right inset).
This indicates that the systemic mechanisms leading to the emergence of degree-correlation have a special signature, which
is not captured when generating assortativity artificially, i.e., {\it ex post facto}.

\begin{figure}
  \centering
\includegraphics[width=\linewidth]{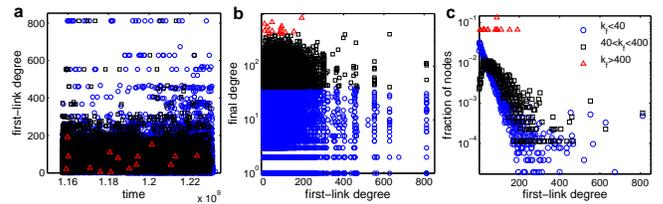}
\caption{\textbf{Nodes' selection mechanisms of their initial neighbors in RWNs.} The Facebook network analyzed in
  Fig.~\ref{fig:realdata}a. (\textbf{a}) Degree of
  the nodes chosen as first connections by those nodes whose final (i.e. at the end of the growth process) 
  degree $k_f$ is low ($k_f<40$, blue circles), high ($k_f>400$, red triangles), and intermediate   ($40\leq k_f \leq 400$, black squares). The
reported values are from the largest connected component of the Facebook network of Fig.~\ref{fig:realdata}a formed only by those
edges that are time-stamped ($N=60,663$, $L=614,541$, $\langle k
\rangle=20$, $r=0.1851$). (\textbf{b}) Log-linear plot of {of the final degree $k_f$ of each node (labeled according to the legend in Fig.~\ref{fig:finalhubs}c) as a function of the degree of its first connection.} (\textbf{c})  {Log-linear plot of the fraction of high (red triangles), medium (black squares) and low (blue circles) degree nodes establishing their first link with a node of a given degree. }\label{fig:finalhubs}}
\end{figure}

Further striking evidence comes to light from {a deeper}
analysis of social RWNs: in some cases the final leaders (i.e. the nodes that, at the end
of the process, do acquire a leading role in terms of their degree)
actually behave {\it anti-preferentially} {when entering the network}. In
Fig.~\ref{fig:finalhubs}, the Facebook network of
Fig.~\ref{fig:realdata}a is examined, and one
sees that{, plotting the degree of the first linked node as a function of time,} those nodes eventually becoming the
network's leaders (i.e. the final hubs, {red triangles}) tend initially (at the moment
at which they start forming part of the network) to link existing nodes with low degree values (Fig.~\ref{fig:finalhubs}a). 
{This is clearer from Fig.~\ref{fig:finalhubs}b where the final degree $k_f$ achieved by a given node, labeled as a red triangle 
($k_f>400$), a black square ($40<k_f<400$) or a blue circle ($k_f<40$), is compared 
to the degree of its first neighbor at the time that node entered the network. 
A straightforward statistical analysis of
the data shows in Fig.~\ref{fig:finalhubs}c that indeed the fraction of final hubs forming initial connections
with nodes of low-medium degrees is far larger than that of the nodes which ultimately acquire intermediate and low degrees.}

\subsection*{The generative model.}
{Following the empircal} observation in Fig.~\ref{fig:finalhubs} {of} a {nexus} between initial anti-preferential
attachments and long-term high degrees,  we propose a generative model which creates SF-like networks  with tunable global assortativity and realistic local assortativity patterns, while also reproduc{ing} the hierarchical structure of the network's clustering.
The model {reflects} a microscopic mechanism for a {\it struggle for leadership} between two competing populations of nodes: type I nodes (acting as {\it followers} and selecting connections so that a preferential attachment rule spontaneously emerges\cite{Boccaletti2007}) and type II nodes (acting as {\it potential leaders}, i.e. adopting  anti-preferential behavior  which leads them to prefer lower degree {nodes} for the establishment of their initial links).

\begin{figure}
  \centering
\includegraphics[width=0.5\linewidth]{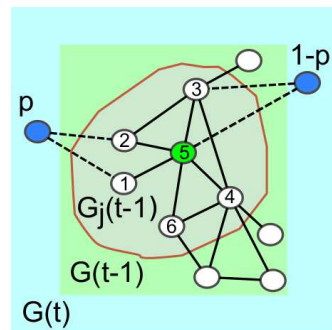}
\caption{\textbf{The network growth process.} At time $t$, the graph $G(t)$ is updated with a
 new node (blue circle) which  forms $m$ connections (in the example $m=2$, dashed
 lines) within the subgraph $G_j(t-1)$ {with a probability $p$ to the
 lowest degree nodes (nodes 1 and 2) or with probability $1-p$ at
 random (nodes 3 and 5). The subgraph $G_j(t-1)$ is composed of a
 randomly chosen  node $j$ (node 5, green circle) and its nearest
 neighbors at time $t-1$. }\label{fig:sketch}}
\end{figure}

Under such a mechanism, a network of $N$ nodes is created by sequentially adding units to an initial clique of $m \le N_0 \lll N$ vertices. The growing
process occurs at discrete times: at each {time} step $1 \leq t \leq N-N_0$ a new node enters the graph, and forms $m$ links 
{with existing nodes} according to an attachment rule that is {illustrated} schematically  in 
Fig.~\ref{fig:sketch} and  summarized as follows:
\begin{enumerate}
\item An anchor node $j$ is selected uniformly at random from the nodes existing at time $t-1$.
\item The subgraph $G_j$  composed of node $j$ and all
  other nodes {that are at distance less than or equal to $\ell$ from
    $j$} {is examined}.
\item With probability $1-p$, the new node behaves as a {\it follower} (type I): it selects $m$ nodes from $G_j$ uniformly at random,
and links to them. With probability $p$, the new node behaves instead as a {\it potential leader} (type II): it forms links with the
$m$ lowest degree nodes in $G_j$.
\end{enumerate}
The parameter $\ell$ is defined as the so called {\em{penetration depth}}, i.e. the extent
of local information (around the anchor $j$) accessible to the entering node. In the following, we set $\ell=1$,
so that ${G}_j$ is the subgraph containing $j$ and all its nearest neighbors. Once $\ell=1$ is set, the model is uniquely determined by two parameters:
the average degree $\langle k\rangle = 2m$ and $p$, the fraction of
type II nodes. In the absence of {\it potential leaders} ($p=0$), the
growth of the resulting network exhibits emergent preferential
attachment and hierarchical clustering{\cite{Boccaletti2007}}: the $p=0$ case produces a pure
SF network with degree distribution {\cite{Barabasi2002,Newman2003b}} $P_k\sim k^{-3}$, and with
additional hierarchical SF clustering{\cite{Ravasz2002}}  $C_k\sim k^{-1}$. This is actually due to the so called {\it
  friendship paradox}{\cite{Scott1991}}, stating that, {averaged across the network}, the neighbors of a
node $i$ will always have {a higher average degree} than $k_i$. 
Since, indeed, the number of subgraphs $G_j$ in which a node $i$ appears is equal to {$k_i+1$}, higher degree nodes will tend to naturally receive more and
more links. It is important to note that this preferential behavior is
in fact, emergent: {the entering nodes do not require global knowledge
of the degree levels in the system, nor any explicit preference for high degree nodes}.  
In that sense, preferential attachment can be viewed as a kind of null behavior {in which the rate of growth increases with size}, 
as the analogous Yule process is understood in evolutionary dynamics\cite{yule-1925,simon-1955}.

\begin{figure}
  \centering
\includegraphics[width=\linewidth]{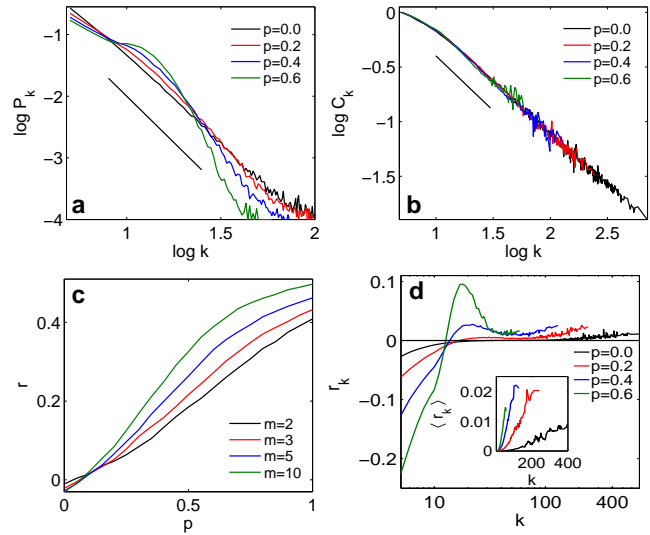}
\caption{\textbf{Emergent topology in the generated network.} (\textbf{a}) Normalized degree distribution $P_k$ (log$_{10}$ scale)
{\it vs.} the logarithm (base 10) of $k$, and (\textbf{b}) $\log_{10}-\log_{10}$ plot of $C_k$ {\it vs.} $k$, for $m=5$ and different values of the probability $p$ (see legend for color-coding).
(\textbf{c}) Assortativity coefficient $r$ {\it vs.} $p$, for different values
of $m$ (see legend for color-coding). (\textbf{d}) Log-linear plot of the local assortativity $r_k$ {(main panel)} and average local assortativity $\langle r_k\rangle$ {(inset)} {\it vs.} $k$, for $m=5$ and several values of
$p$ (see legend for color-coding). 
In all cases, $N=10^4$, $N_0=10$,
and each point refers to an ensemble average over 20 network
realizations. As a guide for the eyes, the straight lines in (\textbf{a}) and
(\textbf{b}) stay for the functions $P_k\propto k^{-3}$ and  $C_k\propto
k^{-1}$, respectively.\label{fig:model} }
\end{figure}

When instead the population is split (with some nodes following the null preferential attachment, and some other{s} linking in an anti-preferential manner), the local assortativity pattern {shown} in Fig.~\ref{fig:realdata}a, characterizing social systems{, emerges}. Namely, the contribution to assortativity from nodes of degree $k$  {\it i)} increases with $k$ from $k=1$ to a local maximum located just above the average degree, {\it ii)} decreases to a subsequent local minimum, and then {\it iii)} increases again as $k\rightarrow\infty$, i.e. qualitatively reproducing the generic tendency {observed} in social RWNs, which is only captured in random generated networks with artificially induced assortativity at the expense of obliterating the graph's clustering traits.
The results {of the model} are summarized in Fig.~\ref{fig:model}. As $p$
increases, the degree distribution of the resulting network deviates
more and more from a pure SF configuration (Fig.~\ref{fig:model}a),
but at the same time the hierarchical clustering traits are entirely
preserved (Fig.~\ref{fig:model}b). The generated network is actually endowed with a fully controllable and tunable level of global assortativity $r$ (as a function of $m$, as shown in Fig.~\ref{fig:model}c), while, more remarkably, the assortativity local pattern is fully reproduced (Fig.~\ref{fig:model}d).

\subsection*{Analytical description.}
We next move toward giving a more analytic description of the motivations and roots 
{underlying} the {proposed} model  and the observed, emergent phenomena.
We start by {noting} that links {in this model} are  undirected, and this  leads to a symmetry of interpretations: one can describe the type II nodes as preferring low-degree units {(as it is described in our generative model)}, 
or one can state that low-degree nodes are more likely to create links with type II newcomers.
The second interpretation is actually in line with what arises from recent sociological studies, 
which indeed indicate that people are limited in the number of relationships 
they can maintain over time (with the exact number of maximal relationships being  an open question).
Starting from the seminal work{s} by  Dunbar\cite{dunbar-1992,hilldunbar-2003}, the limitations on the number 
of active social connections have been {extensively} studied  and empirical support from online social networks has also been adduced\cite{goncalves-2011}. 
{In the present case, the emergence of positive assortativity is associated with the interplay of two mechanisms: } an innate preferential attachment  (resulting from nodes that nonhierarchically form connections with a pre-existing growing structure) 
and a limited ability of human beings to maintain many relationships. 

{
By comparing the average contribution {of assortativity} per node {of degree $k$, $\langle r_k\rangle$,} and the total contribution 
of nodes of degree $k${, $r_k$}, one can actually understand the origin of the peak in the local assortativity.
The average contribution for nodes of degree $k$ increases monotonically with $k$ {(inset of Fig.~\ref{fig:model}d)}.
However, the frequency of nodes decreases monotonically with $k$ in pure scale-free networks {(Fig.~\ref{fig:model}a)}.
With the introduction of type II nodes, lower{-medium} degree nodes become more frequent{, as  observed in Fig.~\ref{fig:model}a for $p=0.6$,} even though an overall scale-free-like degree distribution is maintained. The combination of more-common than expected medium degree nodes and per-node contribution to assortativity that increases with $k$ leads to the characteristic bump observed in the model and the data.}

As the network's growth proceeds, type II nodes actually tend to develop a higher degree on average. 
This is because new links are obtained with probability
\begin{equation}
P \sim \frac{1}{N_t}\frac{m}{\vert G_j\vert}
\end{equation}
where $N_t$ is the number of nodes in the system at time $t$ and $\vert G_j\vert$ is the size of the neighborhood of the subgraph of a given anchor node $j$.
By {choosing anchor} nodes with small $\vert G_j\vert$ (low degree), type II nodes {actually} increase  their likelihood of {being linked from} future, incoming, {nodes}. {Because this increased likelihood can be understood as type II nodes ``placing themselves'' in smaller neighborhoods so that they are more 
likely to be linked to than when chosen at random, 
we understand this advantage as a kind of improved visibility to the linking process}.

\begin{figure}
\centering
\includegraphics[width=\linewidth]{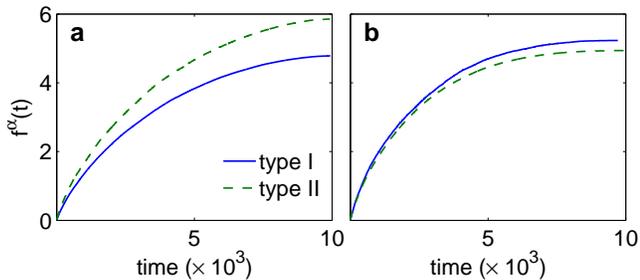}
\caption{\textbf{Emergence of leadership during the growth process.}
Average increased degree (the degree acquired after nodes have first
appeared in the graph, vertical axes) as a function of time
(horizontal axis), for type I (followers) and type II (potential
leaders) nodes, and for (\textbf{a}) $p=0.2$, and (\textbf{b}) $p=0.8$.
See {the Methods section} for the explanation on how the reported values are calculated.
Panels report the average increased degree {$f^{\alpha}(t)$} of the nodes of different
types ($\alpha=$type I or II), after having been in the system for $t$
steps. $N=10^4$, $N_0=5$ and $m=5$. {Color and line style} codes are
defined in the legend of panel (\textbf{a}). \label{fig:degree_evolution}}
\end{figure}

{In fact, one can measure the number of neighbors at time $t$ for each node type as described in the Methods section. } 
The results are shown in Fig.~\ref{fig:degree_evolution}, and point to
the emergence of leadership of type II nodes at low values of $p$
(Fig.~\ref{fig:degree_evolution}a). At intermediate values of $p$
(not shown)  no significant differences
are observed between the two nodes' populations in the way the average
increased degree {evolves} in time. Only at large $p$ values
(Fig.~\ref{fig:degree_evolution}b), where anti-preferential nodes are
vastly predominant in number the trend is actually reversed and type
I nodes (the followers) now seem to be favored in attracting
connections. Such a latter situation corresponds however to a rather
homogeneous network, where a SF-like distribution is no longer
observed (see Fig.~\ref{fig:model} for comparing the large deviations
in the degree distribution {already observed}  at $p=0.6$).


\section*{Discussion}
In summary, assortativity, hierarchical structure and fat-tailed degree distributions (well-approximated by power laws) are structural features manifested 
almost ubiquitously {by } RWNs, and until now no model had linked their emergence with microscopic growing assumptions.
Furthermore, these features have a fundamental role in determining many relevant processes, and/or regulating the network's dynamics and functioning.
Guided by the empirical observation of the growth of the friendship network of Facebook
users, we have shown how the combination of preferential and
anti-preferential attachment mechanisms {acting together} in the  same
generative model (via two distinct node populations), leads to {the
growth of }heterogeneous networks with modified scale-free properties and tunable realistic assortativity, while maintaining the  hierarchical  clustering.
Both our analytical predictions and numerical results indicate that networks constructed in this way match the patterns of local assortativity measured in real-world graphs.
By presenting the first generative model with tunable assortativity, this work sheds new light on the structure and evolution of social networks, {and} counterintuitively suggests that anti-preferential attachment is a mechanism  adopted by a fraction of the nodes during the network's growth, as a strategy for increasing their own leadership.

\section*{Methods}
\subsection*{Local {assortativity/}assortativeness.}
In a network with $N$ nodes, $L$ links and degree distribution $P_k$,
the local  {assortativity} or assortativeness  \cite{piraveenan2008} $r_j$  is defined as the contribution of each node to the network assortativity $r$ and it is calculated as $$r_j={(\alpha_j-\beta_j)}/{\sigma_q^2},$$
with $r_k=\sum_{j|k_j=k} r_j$ being  the total of the local assortativity values
of nodes with a given degree $k$ such that $r=\sum_k r_k$. In the above expression,
$\alpha_j=\frac{\hat{k}_j}{2L}\sum_{i=1}^{k_j} \hat{k}_i,$ and
$\beta_j=k_j\frac{\mu_q^2}{2L}$, being $\hat{k}_j=k_j-1$ the remaining degree of node $j$,
$\hat{k}_i$ the
remaining degrees $\hat{k}_1,\hat{k}_2,\dots,\hat{k}_{k_j}$ of the $k_j$ nodes connected to node $j$, and $\mu_q$ and $\sigma_q$ the first and second moments of the remaining degree distribution
$q_k=\frac{(k+1)P_{k+1}}{\sum_j j P_j}$. 

\subsection*{Measuring the average degree of each node type.}
In order to compare the {average} degree of the two node populations as the model evolves, we label each node uniquely by the step in which it entered the network.
This way, {at time $t$, every node $i$  will have $m$ neighbors with
indices $j < i$, and $k_i(t)-m$ neighbors with indices $j >i$.}
{To compare the degree growth rates of type I and type II nodes, we need to measure the characteristic time for new links to form.}
{To do so, we } consider the set of differences in index values, {$j-i$}, {for each neighbor which linked to $i$ at step $j$}
{\begin{equation}
\tau_i^\alpha = \{(j - i) \,\,|\,\,{\text{with}} j \in {\cal N}_i \land (j>i)\},
\end{equation}}
with $\alpha=I,II$ designating the node type  and ${\cal N}_i\land (j>i)$ the neighborhood of $i$. 
Combining {these sets for all nodes of each type}, one obtains the non-unique set
{\begin{equation}
\tau^{\alpha} = \mathop{\cup}_{i=1}\tau_i^\alpha.
\label{taualpha}
\end{equation}}
{Using} Eq.~(\ref{taualpha}), one can {measure} the expected number of neighbors ({after $t$ steps}) for each node type via
\begin{equation}
f^\alpha(t) = \frac{1}{ N^\alpha} \lvert \{i \,\, |\,\, i\in \tau^\alpha \land i < t\} \rvert,
\end{equation}
where $N^\alpha$ is the total number of nodes of type $\alpha$.
Thus $f^\alpha(t)$ provides the {average} number of {new} neighbors {($\langle k \rangle - m$)}
{that} a node of type $\alpha$ will acquire {after $t$ steps}.

\section*{Acknowledgments }
Work partly supported by the Ministerio de Econom\'ia y Competitividad  
of Spain under projects FIS2012-38949-C03-01 and FIS2013-41057-P. 
I.S.N. acknowledges support from GARECOM, Group of Research Excelence URJC-Banco de Santander. 
S.H. acknowledges support from MULTIPLEX (No. 317532) EU project, the Israel Science Foundation, ONR and DTRA. 
M.M.D. thanks the Azrieli Foundation for the award of an Azrieli Fellowship grant.

\bibliography{references}

\begin{thebibliography}{43}%
\makeatletter
\providecommand \@ifxundefined [1]{%
 \@ifx{#1\undefined}
}%
\providecommand \@ifnum [1]{%
 \ifnum #1\expandafter \@firstoftwo
 \else \expandafter \@secondoftwo
 \fi
}%
\providecommand \@ifx [1]{%
 \ifx #1\expandafter \@firstoftwo
 \else \expandafter \@secondoftwo
 \fi
}%
\providecommand \natexlab [1]{#1}%
\providecommand \enquote  [1]{``#1''}%
\providecommand \bibnamefont  [1]{#1}%
\providecommand \bibfnamefont [1]{#1}%
\providecommand \citenamefont [1]{#1}%
\providecommand \href@noop [0]{\@secondoftwo}%
\providecommand \href [0]{\begingroup \@sanitize@url \@href}%
\providecommand \@href[1]{\@@startlink{#1}\@@href}%
\providecommand \@@href[1]{\endgroup#1\@@endlink}%
\providecommand \@sanitize@url [0]{\catcode `\\12\catcode `\$12\catcode
  `\&12\catcode `\#12\catcode `\^12\catcode `\_12\catcode `\%12\relax}%
\providecommand \@@startlink[1]{}%
\providecommand \@@endlink[0]{}%
\providecommand \url  [0]{\begingroup\@sanitize@url \@url }%
\providecommand \@url [1]{\endgroup\@href {#1}{\urlprefix }}%
\providecommand \urlprefix  [0]{URL }%
\providecommand \Eprint [0]{\href }%
\providecommand \doibase [0]{http://dx.doi.org/}%
\providecommand \selectlanguage [0]{\@gobble}%
\providecommand \bibinfo  [0]{\@secondoftwo}%
\providecommand \bibfield  [0]{\@secondoftwo}%
\providecommand \translation [1]{[#1]}%
\providecommand \BibitemOpen [0]{}%
\providecommand \bibitemStop [0]{}%
\providecommand \bibitemNoStop [0]{.\EOS\space}%
\providecommand \EOS [0]{\spacefactor3000\relax}%
\providecommand \BibitemShut  [1]{\csname bibitem#1\endcsname}%
\let\auto@bib@innerbib\@empty
\bibitem [{\citenamefont {Albert}\ \emph {et~al.}(2000)\citenamefont {Albert},
  \citenamefont {Jeong},\ and\ \citenamefont {Barabasi}}]{Albert2000}%
  \BibitemOpen
  \bibfield  {author} {\bibinfo {author} {\bibfnamefont {R.}~\bibnamefont
  {Albert}}, \bibinfo {author} {\bibfnamefont {H.}~\bibnamefont {Jeong}}, \
  and\ \bibinfo {author} {\bibfnamefont {A.-L.}\ \bibnamefont {Barabasi}},\
  }\href@noop {} {\bibfield  {journal} {\bibinfo  {journal} {Nature}\ }\textbf
  {\bibinfo {volume} {406}},\ \bibinfo {pages} {378} (\bibinfo {year}
  {2000})}\BibitemShut {NoStop}%
\bibitem [{\citenamefont {Albert}\ and\ \citenamefont
  {Barab\'asi}(2002)}]{Barabasi2002}%
  \BibitemOpen
  \bibfield  {author} {\bibinfo {author} {\bibfnamefont {R.}~\bibnamefont
  {Albert}}\ and\ \bibinfo {author} {\bibfnamefont {A.-L.}\ \bibnamefont
  {Barab\'asi}},\ }\href {\doibase 10.1103/RevModPhys.74.47} {\bibfield
  {journal} {\bibinfo  {journal} {Rev. Mod. Phys}\ }\textbf {\bibinfo {volume}
  {74}},\ \bibinfo {pages} {47} (\bibinfo {year} {2002})}\BibitemShut {NoStop}%
\bibitem [{\citenamefont {Wuchty}\ \emph {et~al.}(2003)\citenamefont {Wuchty},
  \citenamefont {Oltvai},\ and\ \citenamefont {Barabasi}}]{Wuchty2003}%
  \BibitemOpen
  \bibfield  {author} {\bibinfo {author} {\bibfnamefont {S.}~\bibnamefont
  {Wuchty}}, \bibinfo {author} {\bibfnamefont {Z.~N.}\ \bibnamefont {Oltvai}},
  \ and\ \bibinfo {author} {\bibfnamefont {A.-L.}\ \bibnamefont {Barabasi}},\
  }\href@noop {} {\bibfield  {journal} {\bibinfo  {journal} {Nat. Genet.}\
  }\textbf {\bibinfo {volume} {35}},\ \bibinfo {pages} {176} (\bibinfo {year}
  {2003})}\BibitemShut {NoStop}%
\bibitem [{\citenamefont {Newman}(2003{\natexlab{a}})}]{Newman2003b}%
  \BibitemOpen
  \bibfield  {author} {\bibinfo {author} {\bibfnamefont {M.~E.~J.}\
  \bibnamefont {Newman}},\ }\href {\doibase 10.1137/S003614450342480}
  {\bibfield  {journal} {\bibinfo  {journal} {SIAM Review}\ }\textbf {\bibinfo
  {volume} {45}},\ \bibinfo {pages} {167} (\bibinfo {year}
  {2003}{\natexlab{a}})}\BibitemShut {NoStop}%
\bibitem [{\citenamefont {Boccaletti}\ \emph {et~al.}(2006)\citenamefont
  {Boccaletti}, \citenamefont {Latora}, \citenamefont {Moreno}, \citenamefont
  {Chavez},\ and\ \citenamefont {Hwang}}]{Boccaletti2006}%
  \BibitemOpen
  \bibfield  {author} {\bibinfo {author} {\bibfnamefont {S.}~\bibnamefont
  {Boccaletti}}, \bibinfo {author} {\bibfnamefont {V.}~\bibnamefont {Latora}},
  \bibinfo {author} {\bibfnamefont {Y.}~\bibnamefont {Moreno}}, \bibinfo
  {author} {\bibfnamefont {M.}~\bibnamefont {Chavez}}, \ and\ \bibinfo {author}
  {\bibfnamefont {D.}~\bibnamefont {Hwang}},\ }\href@noop {} {\bibfield
  {journal} {\bibinfo  {journal} {Phys. Rep.}\ }\textbf {\bibinfo {volume}
  {424}},\ \bibinfo {pages} {175} (\bibinfo {year} {2006})}\BibitemShut
  {NoStop}%
\bibitem [{\citenamefont {Boccaletti}\ \emph {et~al.}(2014)\citenamefont
  {Boccaletti}, \citenamefont {Bianconi}, \citenamefont {Criado}, \citenamefont
  {del Genio}, \citenamefont {G\'{o}mez-Garde\~{n}es}, \citenamefont {Romance},
  \citenamefont {Sendi{\~n}a-Nadal}, \citenamefont {Wang},\ and\ \citenamefont
  {Zanin}}]{Boccaletti2014}%
  \BibitemOpen
  \bibfield  {author} {\bibinfo {author} {\bibfnamefont {S.}~\bibnamefont
  {Boccaletti}}, \bibinfo {author} {\bibfnamefont {G.}~\bibnamefont
  {Bianconi}}, \bibinfo {author} {\bibfnamefont {R.}~\bibnamefont {Criado}},
  \bibinfo {author} {\bibfnamefont {C.~I.}\ \bibnamefont {del Genio}}, \bibinfo
  {author} {\bibfnamefont {J.}~\bibnamefont {G\'{o}mez-Garde\~{n}es}}, \bibinfo
  {author} {\bibfnamefont {M.}~\bibnamefont {Romance}}, \bibinfo {author}
  {\bibfnamefont {I.}~\bibnamefont {Sendi{\~n}a-Nadal}}, \bibinfo {author}
  {\bibfnamefont {Z.}~\bibnamefont {Wang}}, \ and\ \bibinfo {author}
  {\bibfnamefont {M.}~\bibnamefont {Zanin}},\ }\href {\doibase
  10.1016/j.physrep.2014.07.001} {\bibfield  {journal} {\bibinfo  {journal}
  {Phys. Rep.}\ }\textbf {\bibinfo {volume} {544}},\ \bibinfo {pages} {1}
  (\bibinfo {year} {2014})}\BibitemShut {NoStop}%
\bibitem [{\citenamefont {Ravasz}\ \emph {et~al.}(2002)\citenamefont {Ravasz},
  \citenamefont {Somera}, \citenamefont {Mongru}, \citenamefont {Oltvai},\ and\
  \citenamefont {Barab\'{a}si}}]{Ravasz2002}%
  \BibitemOpen
  \bibfield  {author} {\bibinfo {author} {\bibfnamefont {E.}~\bibnamefont
  {Ravasz}}, \bibinfo {author} {\bibfnamefont {A.~L.}\ \bibnamefont {Somera}},
  \bibinfo {author} {\bibfnamefont {D.~A.}\ \bibnamefont {Mongru}}, \bibinfo
  {author} {\bibfnamefont {Z.~N.}\ \bibnamefont {Oltvai}}, \ and\ \bibinfo
  {author} {\bibfnamefont {A.~L.}\ \bibnamefont {Barab\'{a}si}},\ }\href
  {\doibase 10.1126/science.1073374} {\bibfield  {journal} {\bibinfo  {journal}
  {Science}\ }\textbf {\bibinfo {volume} {297}},\ \bibinfo {pages} {1551}
  (\bibinfo {year} {2002})}\BibitemShut {NoStop}%
\bibitem [{\citenamefont {Newman}(2002)}]{Newman2002}%
  \BibitemOpen
  \bibfield  {author} {\bibinfo {author} {\bibfnamefont {M.}~\bibnamefont
  {Newman}},\ }\href {\doibase 10.1103/PhysRevLett.89.208701} {\bibfield
  {journal} {\bibinfo  {journal} {Phys. Rev. Lett.}\ }\textbf {\bibinfo
  {volume} {89}},\ \bibinfo {pages} {208701} (\bibinfo {year}
  {2002})}\BibitemShut {NoStop}%
\bibitem [{\citenamefont {Newman}(2003{\natexlab{b}})}]{Newman2003}%
  \BibitemOpen
  \bibfield  {author} {\bibinfo {author} {\bibfnamefont {M.}~\bibnamefont
  {Newman}},\ }\href {\doibase 10.1103/PhysRevE.67.026126} {\bibfield
  {journal} {\bibinfo  {journal} {Phys. Rev. E}\ }\textbf {\bibinfo {volume}
  {67}},\ \bibinfo {pages} {026126} (\bibinfo {year}
  {2003}{\natexlab{b}})}\BibitemShut {NoStop}%
\bibitem [{\citenamefont {Boccaletti}\ \emph {et~al.}(2007)\citenamefont
  {Boccaletti}, \citenamefont {Hwang},\ and\ \citenamefont
  {Latora}}]{Boccaletti2007}%
  \BibitemOpen
  \bibfield  {author} {\bibinfo {author} {\bibfnamefont {S.}~\bibnamefont
  {Boccaletti}}, \bibinfo {author} {\bibfnamefont {D.-U.}\ \bibnamefont
  {Hwang}}, \ and\ \bibinfo {author} {\bibfnamefont {V.}~\bibnamefont
  {Latora}},\ }\href {\doibase 10.1142/S0218127407018518} {\bibfield  {journal}
  {\bibinfo  {journal} {IJBC}\ }\textbf {\bibinfo {volume} {17}},\ \bibinfo
  {pages} {2447} (\bibinfo {year} {2007})}\BibitemShut {NoStop}%
\bibitem [{\citenamefont {Brede}(2011)}]{Brede2011}%
  \BibitemOpen
  \bibfield  {author} {\bibinfo {author} {\bibfnamefont {M.}~\bibnamefont
  {Brede}},\ }\href {\doibase 10.1162/artl_a_00039} {\bibfield  {journal}
  {\bibinfo  {journal} {Artificial Life}\ }\textbf {\bibinfo {volume} {17}},\
  \bibinfo {pages} {281} (\bibinfo {year} {2011})}\BibitemShut {NoStop}%
\bibitem [{\citenamefont {Barab\'asi}\ and\ \citenamefont
  {Albert}(1999)}]{Barabasi1999}%
  \BibitemOpen
  \bibfield  {author} {\bibinfo {author} {\bibfnamefont {A.-L.}\ \bibnamefont
  {Barab\'asi}}\ and\ \bibinfo {author} {\bibfnamefont {R.~a.}\ \bibnamefont
  {Albert}},\ }\href {\doibase 10.1126/science.286.5439.509} {\bibfield
  {journal} {\bibinfo  {journal} {Science}\ }\textbf {\bibinfo {volume}
  {286}},\ \bibinfo {pages} {509} (\bibinfo {year} {1999})}\BibitemShut
  {NoStop}%
\bibitem [{\citenamefont {Dorogovtsev}\ \emph {et~al.}(2000)\citenamefont
  {Dorogovtsev}, \citenamefont {Mendes},\ and\ \citenamefont
  {Samukhin}}]{Dorogovtsev2000}%
  \BibitemOpen
  \bibfield  {author} {\bibinfo {author} {\bibfnamefont {S.~N.}\ \bibnamefont
  {Dorogovtsev}}, \bibinfo {author} {\bibfnamefont {J.~F.}\ \bibnamefont
  {Mendes}}, \ and\ \bibinfo {author} {\bibfnamefont {a.~N.}\ \bibnamefont
  {Samukhin}},\ }\href@noop {} {\bibfield  {journal} {\bibinfo  {journal}
  {Phys. Rev. Lett.}\ }\textbf {\bibinfo {volume} {85}},\ \bibinfo {pages}
  {4633} (\bibinfo {year} {2000})}\BibitemShut {NoStop}%
\bibitem [{\citenamefont {V\'azquez}(2003)}]{Vazquez2003}%
  \BibitemOpen
  \bibfield  {author} {\bibinfo {author} {\bibfnamefont {A.}~\bibnamefont
  {V\'azquez}},\ }\href {\doibase 10.1103/PhysRevE.67.056104} {\bibfield
  {journal} {\bibinfo  {journal} {Phys. Rev. E}\ }\textbf {\bibinfo {volume}
  {67}},\ \bibinfo {pages} {056104} (\bibinfo {year} {2003})}\BibitemShut
  {NoStop}%
\bibitem [{\citenamefont {Chung}\ \emph {et~al.}(2003)\citenamefont {Chung},
  \citenamefont {Lu}, \citenamefont {Dewey},\ and\ \citenamefont
  {Galas}}]{Chung2003}%
  \BibitemOpen
  \bibfield  {author} {\bibinfo {author} {\bibfnamefont {F.}~\bibnamefont
  {Chung}}, \bibinfo {author} {\bibfnamefont {L.}~\bibnamefont {Lu}}, \bibinfo
  {author} {\bibfnamefont {T.~G.}\ \bibnamefont {Dewey}}, \ and\ \bibinfo
  {author} {\bibfnamefont {D.~J.}\ \bibnamefont {Galas}},\ }\href {\doibase
  doi:10.1089/106652703322539024} {\bibfield  {journal} {\bibinfo  {journal}
  {J. Comp. Biol.}\ ,\ \bibinfo {pages} {677}} (\bibinfo {year}
  {2003})}\BibitemShut {NoStop}%
\bibitem [{\citenamefont {Krapivsky}\ and\ \citenamefont
  {Redner}(2005)}]{Krapivsky2005}%
  \BibitemOpen
  \bibfield  {author} {\bibinfo {author} {\bibfnamefont {P.~L.}\ \bibnamefont
  {Krapivsky}}\ and\ \bibinfo {author} {\bibfnamefont {S.}~\bibnamefont
  {Redner}},\ }\href {\doibase 10.1103/PhysRevE.71.036118} {\bibfield
  {journal} {\bibinfo  {journal} {Phys. Rev. E}\ }\textbf {\bibinfo {volume}
  {71}},\ \bibinfo {pages} {036118} (\bibinfo {year} {2005})}\BibitemShut
  {NoStop}%
\bibitem [{\citenamefont {Clauset}\ \emph {et~al.}(2008)\citenamefont
  {Clauset}, \citenamefont {Moore},\ and\ \citenamefont
  {Newman}}]{Clauset2008}%
  \BibitemOpen
  \bibfield  {author} {\bibinfo {author} {\bibfnamefont {A.}~\bibnamefont
  {Clauset}}, \bibinfo {author} {\bibfnamefont {C.}~\bibnamefont {Moore}}, \
  and\ \bibinfo {author} {\bibfnamefont {M.~E.~J.}\ \bibnamefont {Newman}},\
  }\href@noop {} {\bibfield  {journal} {\bibinfo  {journal} {Nature}\ }\textbf
  {\bibinfo {volume} {453}},\ \bibinfo {pages} {98} (\bibinfo {year}
  {2008})}\BibitemShut {NoStop}%
\bibitem [{\citenamefont {Lorimer}\ \emph {et~al.}(2015)\citenamefont
  {Lorimer}, \citenamefont {Gomez},\ and\ \citenamefont {Stoop}}]{Lorimer2015}%
  \BibitemOpen
  \bibfield  {author} {\bibinfo {author} {\bibfnamefont {T.}~\bibnamefont
  {Lorimer}}, \bibinfo {author} {\bibfnamefont {F.}~\bibnamefont {Gomez}}, \
  and\ \bibinfo {author} {\bibfnamefont {R.}~\bibnamefont {Stoop}},\ }\href
  {\doibase 10.1038/srep12353} {\bibfield  {journal} {\bibinfo  {journal} {Sci.
  Rep.}\ }\textbf {\bibinfo {volume} {5}},\ \bibinfo {pages} {12353} (\bibinfo
  {year} {2015})}\BibitemShut {NoStop}%
\bibitem [{\citenamefont {Bender}\ and\ \citenamefont
  {Canfield}(1978)}]{Bender1978}%
  \BibitemOpen
  \bibfield  {author} {\bibinfo {author} {\bibfnamefont {E.~A.}\ \bibnamefont
  {Bender}}\ and\ \bibinfo {author} {\bibfnamefont {E.}~\bibnamefont
  {Canfield}},\ }\href {\doibase
  http://dx.doi.org/10.1016/0097-3165(78)90059-6} {\bibfield  {journal}
  {\bibinfo  {journal} {Journal of Combinatorial Theory, Series A}\ }\textbf
  {\bibinfo {volume} {24}},\ \bibinfo {pages} {296 } (\bibinfo {year}
  {1978})}\BibitemShut {NoStop}%
\bibitem [{\citenamefont {Zhou}\ \emph {et~al.}(2008)\citenamefont {Zhou},
  \citenamefont {Xu}, \citenamefont {Zhang}, \citenamefont {Sun}, \citenamefont
  {Small},\ and\ \citenamefont {Lu}}]{Zhou2008}%
  \BibitemOpen
  \bibfield  {author} {\bibinfo {author} {\bibfnamefont {J.}~\bibnamefont
  {Zhou}}, \bibinfo {author} {\bibfnamefont {X.}~\bibnamefont {Xu}}, \bibinfo
  {author} {\bibfnamefont {J.}~\bibnamefont {Zhang}}, \bibinfo {author}
  {\bibfnamefont {J.}~\bibnamefont {Sun}}, \bibinfo {author} {\bibfnamefont
  {M.}~\bibnamefont {Small}}, \ and\ \bibinfo {author} {\bibfnamefont {J.-a.}\
  \bibnamefont {Lu}},\ }\href {\doibase 10.1142/S0218127408022536} {\bibfield
  {journal} {\bibinfo  {journal} {IJBC}\ }\textbf {\bibinfo {volume} {18}},\
  \bibinfo {pages} {3495} (\bibinfo {year} {2008})}\BibitemShut {NoStop}%
\bibitem [{\citenamefont {Bassler}\ \emph {et~al.}(2015)\citenamefont
  {Bassler}, \citenamefont {Genio}, \citenamefont {Erd\"os}, \citenamefont
  {Mikl\'{o}s},\ and\ \citenamefont {Toroczkai}}]{Bassler2015}%
  \BibitemOpen
  \bibfield  {author} {\bibinfo {author} {\bibfnamefont {K.~E.}\ \bibnamefont
  {Bassler}}, \bibinfo {author} {\bibfnamefont {C.~I.~D.}\ \bibnamefont
  {Genio}}, \bibinfo {author} {\bibfnamefont {P.~L.}\ \bibnamefont {Erd\"os}},
  \bibinfo {author} {\bibfnamefont {I.}~\bibnamefont {Mikl\'{o}s}}, \ and\
  \bibinfo {author} {\bibfnamefont {Z.}~\bibnamefont {Toroczkai}},\ }\href
  {\doibase 10.1088/1367-2630/17/8/083052} {\bibfield  {journal} {\bibinfo
  {journal} {New J. Phys.}\ }\textbf {\bibinfo {volume} {17}},\ \bibinfo
  {pages} {083052} (\bibinfo {year} {2015})}\BibitemShut {NoStop}%
\bibitem [{\citenamefont {Catanzaro}\ \emph {et~al.}(2004)\citenamefont
  {Catanzaro}, \citenamefont {Caldarelli},\ and\ \citenamefont
  {Pietronero}}]{Catanzaro2004}%
  \BibitemOpen
  \bibfield  {author} {\bibinfo {author} {\bibfnamefont {M.}~\bibnamefont
  {Catanzaro}}, \bibinfo {author} {\bibfnamefont {G.}~\bibnamefont
  {Caldarelli}}, \ and\ \bibinfo {author} {\bibfnamefont {L.}~\bibnamefont
  {Pietronero}},\ }\href {\doibase 10.1016/j.physa.2004.02.033} {\bibfield
  {journal} {\bibinfo  {journal} {Physica A}\ }\textbf {\bibinfo {volume}
  {338}},\ \bibinfo {pages} {119} (\bibinfo {year} {2004})}\BibitemShut
  {NoStop}%
\bibitem [{\citenamefont {Quayle}\ \emph {et~al.}(2006)\citenamefont {Quayle},
  \citenamefont {Siddiqui},\ and\ \citenamefont {Jones}}]{Quayle2006}%
  \BibitemOpen
  \bibfield  {author} {\bibinfo {author} {\bibfnamefont {A.}~\bibnamefont
  {Quayle}}, \bibinfo {author} {\bibfnamefont {A.}~\bibnamefont {Siddiqui}}, \
  and\ \bibinfo {author} {\bibfnamefont {S.~J.}\ \bibnamefont {Jones}},\ }\href
  {\doibase 10.1140/epjb/e2006-00170-5} {\bibfield  {journal} {\bibinfo
  {journal} {EPJ B}\ }\textbf {\bibinfo {volume} {50}},\ \bibinfo {pages} {617}
  (\bibinfo {year} {2006})}\BibitemShut {NoStop}%
\bibitem [{\citenamefont {Toivonen}\ \emph {et~al.}(2006)\citenamefont
  {Toivonen}, \citenamefont {Onnela}, \citenamefont {Saram\"{a}ki},
  \citenamefont {Hyv\"{o}nen},\ and\ \citenamefont {Kaski}}]{Toivonen2006}%
  \BibitemOpen
  \bibfield  {author} {\bibinfo {author} {\bibfnamefont {R.}~\bibnamefont
  {Toivonen}}, \bibinfo {author} {\bibfnamefont {J.-P.}\ \bibnamefont
  {Onnela}}, \bibinfo {author} {\bibfnamefont {J.}~\bibnamefont
  {Saram\"{a}ki}}, \bibinfo {author} {\bibfnamefont {J.}~\bibnamefont
  {Hyv\"{o}nen}}, \ and\ \bibinfo {author} {\bibfnamefont {K.}~\bibnamefont
  {Kaski}},\ }\href {\doibase 10.1016/j.physa.2006.03.050} {\bibfield
  {journal} {\bibinfo  {journal} {Physica A}\ }\textbf {\bibinfo {volume}
  {371}},\ \bibinfo {pages} {851} (\bibinfo {year} {2006})}\BibitemShut
  {NoStop}%
\bibitem [{\citenamefont {Nguyen}\ and\ \citenamefont {Tran}(2010)}]{Tran2010}%
  \BibitemOpen
  \bibfield  {author} {\bibinfo {author} {\bibfnamefont {K.}~\bibnamefont
  {Nguyen}}\ and\ \bibinfo {author} {\bibfnamefont {D.}~\bibnamefont {Tran}},\
  }in\ \href {\doibase 10.1109/ICCE.2010.5670676} {\emph {\bibinfo {booktitle}
  {Communications and Electronics (ICCE)}}}\ (\bibinfo {year} {2010})\ pp.\
  \bibinfo {pages} {30--35}\BibitemShut {NoStop}%
\bibitem [{\citenamefont {Shang}(2014)}]{Shang2014}%
  \BibitemOpen
  \bibfield  {author} {\bibinfo {author} {\bibfnamefont {Y.}~\bibnamefont
  {Shang}},\ }\href {\doibase 10.1155/2014/759391} {\bibfield  {journal}
  {\bibinfo  {journal} {The Scientific World Journal}\ }\textbf {\bibinfo
  {volume} {2014}},\ \bibinfo {pages} {759391} (\bibinfo {year}
  {2014})}\BibitemShut {NoStop}%
\bibitem [{\citenamefont {\v{S}ubelj}\ and\ \citenamefont
  {Bajec}(2013)}]{Subelj2013}%
  \BibitemOpen
  \bibfield  {author} {\bibinfo {author} {\bibfnamefont {L.}~\bibnamefont
  {\v{S}ubelj}}\ and\ \bibinfo {author} {\bibfnamefont {M.}~\bibnamefont
  {Bajec}},\ }in\ \href@noop {} {\emph {\bibinfo {booktitle} {22Nd
  International Conference on WWW Companion}}}\ (\bibinfo {year} {2013})\ pp.\
  \bibinfo {pages} {527--530}\BibitemShut {NoStop}%
\bibitem [{\citenamefont {Kunegis}(2013)}]{konect}%
  \BibitemOpen
  \bibfield  {author} {\bibinfo {author} {\bibfnamefont {J.}~\bibnamefont
  {Kunegis}},\ }in\ \href {http://konect.uni-koblenz.de/} {\emph {\bibinfo
  {booktitle} {Int. Web Observatory Workshop}}}\ (\bibinfo {year} {2013})\ pp.\
  \bibinfo {pages} {1343--1350}\BibitemShut {NoStop}%
\bibitem [{\citenamefont {Viswanath}\ \emph {et~al.}(2009)\citenamefont
  {Viswanath}, \citenamefont {Mislove}, \citenamefont {Cha},\ and\
  \citenamefont {Gummadi}}]{facebook-data}%
  \BibitemOpen
  \bibfield  {author} {\bibinfo {author} {\bibfnamefont {B.}~\bibnamefont
  {Viswanath}}, \bibinfo {author} {\bibfnamefont {A.}~\bibnamefont {Mislove}},
  \bibinfo {author} {\bibfnamefont {M.}~\bibnamefont {Cha}}, \ and\ \bibinfo
  {author} {\bibfnamefont {K.~P.}\ \bibnamefont {Gummadi}},\ }in\ \href
  {\doibase 10.1145/1592665.1592675} {\emph {\bibinfo {booktitle} {2nd ACM
  Workshop on Online Social Networks}}},\ \bibinfo {series and number} {WOSN
  '09}\ (\bibinfo  {publisher} {ACM},\ \bibinfo {year} {2009})\ pp.\ \bibinfo
  {pages} {37--42}\BibitemShut {NoStop}%
\bibitem [{\citenamefont {Teller}\ \emph {et~al.}(2014)\citenamefont {Teller},
  \citenamefont {Granell}, \citenamefont {{De Domenico}}, \citenamefont
  {Soriano}, \citenamefont {Gomez},\ and\ \citenamefont {Arenas}}]{Teller2014}%
  \BibitemOpen
  \bibfield  {author} {\bibinfo {author} {\bibfnamefont {S.}~\bibnamefont
  {Teller}}, \bibinfo {author} {\bibfnamefont {C.}~\bibnamefont {Granell}},
  \bibinfo {author} {\bibfnamefont {M.}~\bibnamefont {{De Domenico}}}, \bibinfo
  {author} {\bibfnamefont {J.}~\bibnamefont {Soriano}}, \bibinfo {author}
  {\bibfnamefont {S.}~\bibnamefont {Gomez}}, \ and\ \bibinfo {author}
  {\bibfnamefont {A.}~\bibnamefont {Arenas}},\ }\href {\doibase
  10.1371/journal.pcbi.1003796} {\bibfield  {journal} {\bibinfo  {journal}
  {PLoS Comput. Biol.}\ }\textbf {\bibinfo {volume} {10}},\ \bibinfo {pages}
  {e1003796} (\bibinfo {year} {2014})}\BibitemShut {NoStop}%
\bibitem [{\citenamefont {de~Santos-Sierra}\ \emph {et~al.}(2014)\citenamefont
  {de~Santos-Sierra}, \citenamefont {Sendi\~{n}a Nadal}, \citenamefont {Leyva},
  \citenamefont {Almendral}, \citenamefont {Anava}, \citenamefont {Ayali},
  \citenamefont {Papo},\ and\ \citenamefont {Boccaletti}}]{deSantos2014}%
  \BibitemOpen
  \bibfield  {author} {\bibinfo {author} {\bibfnamefont {D.}~\bibnamefont
  {de~Santos-Sierra}}, \bibinfo {author} {\bibfnamefont {I.}~\bibnamefont
  {Sendi\~{n}a Nadal}}, \bibinfo {author} {\bibfnamefont {I.}~\bibnamefont
  {Leyva}}, \bibinfo {author} {\bibfnamefont {J.~a.}\ \bibnamefont
  {Almendral}}, \bibinfo {author} {\bibfnamefont {S.}~\bibnamefont {Anava}},
  \bibinfo {author} {\bibfnamefont {A.}~\bibnamefont {Ayali}}, \bibinfo
  {author} {\bibfnamefont {D.}~\bibnamefont {Papo}}, \ and\ \bibinfo {author}
  {\bibfnamefont {S.}~\bibnamefont {Boccaletti}},\ }\href {\doibase
  10.1371/journal.pone.0085828} {\bibfield  {journal} {\bibinfo  {journal}
  {PLoS ONE}\ }\textbf {\bibinfo {volume} {9}},\ \bibinfo {pages} {e85828}
  (\bibinfo {year} {2014})}\BibitemShut {NoStop}%
\bibitem [{\citenamefont {Piraveenan}\ \emph {et~al.}(2008)\citenamefont
  {Piraveenan}, \citenamefont {Prokopenko},\ and\ \citenamefont
  {Zomaya}}]{piraveenan2008}%
  \BibitemOpen
  \bibfield  {author} {\bibinfo {author} {\bibfnamefont {M.}~\bibnamefont
  {Piraveenan}}, \bibinfo {author} {\bibfnamefont {M.}~\bibnamefont
  {Prokopenko}}, \ and\ \bibinfo {author} {\bibfnamefont {a.~Y.}\ \bibnamefont
  {Zomaya}},\ }\href {\doibase 10.1209/0295-5075/84/28002} {\bibfield
  {journal} {\bibinfo  {journal} {EPL}\ }\textbf {\bibinfo {volume} {84}},\
  \bibinfo {pages} {28002} (\bibinfo {year} {2008})}\BibitemShut {NoStop}%
\bibitem [{\citenamefont {Leskovec}\ \emph {et~al.}(2007)\citenamefont
  {Leskovec}, \citenamefont {Kleinberg},\ and\ \citenamefont
  {Faloutsos}}]{konect:leskovec107}%
  \BibitemOpen
  \bibfield  {author} {\bibinfo {author} {\bibfnamefont {J.}~\bibnamefont
  {Leskovec}}, \bibinfo {author} {\bibfnamefont {J.}~\bibnamefont {Kleinberg}},
  \ and\ \bibinfo {author} {\bibfnamefont {C.}~\bibnamefont {Faloutsos}},\
  }\href@noop {} {\bibfield  {journal} {\bibinfo  {journal} {ACM Trans.
  Knowledge Discovery from Data}\ }\textbf {\bibinfo {volume} {1}},\ \bibinfo
  {pages} {1} (\bibinfo {year} {2007})}\BibitemShut {NoStop}%
\bibitem [{\citenamefont {Whitfield}(2008)}]{Whitfield2008}%
  \BibitemOpen
  \bibfield  {author} {\bibinfo {author} {\bibfnamefont {J.}~\bibnamefont
  {Whitfield}},\ }\href {\doibase 10.1038/455720a} {\bibfield  {journal}
  {\bibinfo  {journal} {Nature}\ }\textbf {\bibinfo {volume} {455}},\ \bibinfo
  {pages} {720} (\bibinfo {year} {2008})}\BibitemShut {NoStop}%
\bibitem [{\citenamefont {Eom}\ and\ \citenamefont {Jo}(2014)}]{Eom2014}%
  \BibitemOpen
  \bibfield  {author} {\bibinfo {author} {\bibfnamefont {Y.-H.}\ \bibnamefont
  {Eom}}\ and\ \bibinfo {author} {\bibfnamefont {H.-H.}\ \bibnamefont {Jo}},\
  }\href {\doibase 10.1038/srep04603} {\bibfield  {journal} {\bibinfo
  {journal} {Sci. Rep.}\ }\textbf {\bibinfo {volume} {4}},\ \bibinfo {pages}
  {4603} (\bibinfo {year} {2014})}\BibitemShut {NoStop}%
\bibitem [{\citenamefont {Colizza}\ \emph {et~al.}(2007)\citenamefont
  {Colizza}, \citenamefont {Pastor-Satorras},\ and\ \citenamefont
  {Vespignani}}]{Colizza2007}%
  \BibitemOpen
  \bibfield  {author} {\bibinfo {author} {\bibfnamefont {V.}~\bibnamefont
  {Colizza}}, \bibinfo {author} {\bibfnamefont {R.}~\bibnamefont
  {Pastor-Satorras}}, \ and\ \bibinfo {author} {\bibfnamefont {A.}~\bibnamefont
  {Vespignani}},\ }\href@noop {} {\bibfield  {journal} {\bibinfo  {journal}
  {Nat. Phys.}\ }\textbf {\bibinfo {volume} {3}},\ \bibinfo {pages} {276}
  (\bibinfo {year} {2007})}\BibitemShut {NoStop}%
\bibitem [{\citenamefont {Xulvi-Brunet}\ and\ \citenamefont
  {Sokolov}(2004)}]{Xulvi-Brunet2004}%
  \BibitemOpen
  \bibfield  {author} {\bibinfo {author} {\bibfnamefont {R.}~\bibnamefont
  {Xulvi-Brunet}}\ and\ \bibinfo {author} {\bibfnamefont {I.}~\bibnamefont
  {Sokolov}},\ }\href {\doibase 10.1103/PhysRevE.70.066102} {\bibfield
  {journal} {\bibinfo  {journal} {Phys. Rev. E}\ }\textbf {\bibinfo {volume}
  {70}},\ \bibinfo {pages} {066102} (\bibinfo {year} {2004})}\BibitemShut
  {NoStop}%
\bibitem [{\citenamefont {Feld}(1991)}]{Scott1991}%
  \BibitemOpen
  \bibfield  {author} {\bibinfo {author} {\bibfnamefont {S.~L.}\ \bibnamefont
  {Feld}},\ }\href@noop {} {\bibfield  {journal} {\bibinfo  {journal} {American
  Journal of Sociology}\ }\textbf {\bibinfo {volume} {96}},\ \bibinfo {pages}
  {1464} (\bibinfo {year} {1991})}\BibitemShut {NoStop}%
\bibitem [{\citenamefont {Yule}(1925)}]{yule-1925}%
  \BibitemOpen
  \bibfield  {author} {\bibinfo {author} {\bibfnamefont {G.~U.}\ \bibnamefont
  {Yule}},\ }\href {\doibase 10.1098/rstb.1925.0002} {\bibfield  {journal}
  {\bibinfo  {journal} {Phil. Trans. R. Soc. B}\ }\textbf {\bibinfo {volume}
  {213}},\ \bibinfo {pages} {21} (\bibinfo {year} {1925})}\BibitemShut
  {NoStop}%
\bibitem [{\citenamefont {Simon}(1955)}]{simon-1955}%
  \BibitemOpen
  \bibfield  {author} {\bibinfo {author} {\bibfnamefont {H.~A.}\ \bibnamefont
  {Simon}},\ }\href {\doibase 10.1093/biomet/42.3-4.425} {\bibfield  {journal}
  {\bibinfo  {journal} {Biometrika}\ }\textbf {\bibinfo {volume} {42}},\
  \bibinfo {pages} {425} (\bibinfo {year} {1955})}\BibitemShut {NoStop}%
\bibitem [{\citenamefont {Dunbar}(1992)}]{dunbar-1992}%
  \BibitemOpen
  \bibfield  {author} {\bibinfo {author} {\bibfnamefont {R.~I.}\ \bibnamefont
  {Dunbar}},\ }\href@noop {} {\bibfield  {journal} {\bibinfo  {journal}
  {Journal of Human Evolution}\ }\textbf {\bibinfo {volume} {22}},\ \bibinfo
  {pages} {469} (\bibinfo {year} {1992})}\BibitemShut {NoStop}%
\bibitem [{\citenamefont {Hill}\ and\ \citenamefont
  {Dunbar}(2003)}]{hilldunbar-2003}%
  \BibitemOpen
  \bibfield  {author} {\bibinfo {author} {\bibfnamefont {R.}~\bibnamefont
  {Hill}}\ and\ \bibinfo {author} {\bibfnamefont {R.}~\bibnamefont {Dunbar}},\
  }\href {\doibase 10.1007/s12110-003-1016-y} {\bibfield  {journal} {\bibinfo
  {journal} {Human Nature}\ }\textbf {\bibinfo {volume} {14}},\ \bibinfo
  {pages} {53} (\bibinfo {year} {2003})}\BibitemShut {NoStop}%
\bibitem [{\citenamefont {Gon\c{c}alves}\ \emph {et~al.}(2011)\citenamefont
  {Gon\c{c}alves}, \citenamefont {Perra},\ and\ \citenamefont
  {Vespignani}}]{goncalves-2011}%
  \BibitemOpen
  \bibfield  {author} {\bibinfo {author} {\bibfnamefont {B.}~\bibnamefont
  {Gon\c{c}alves}}, \bibinfo {author} {\bibfnamefont {N.}~\bibnamefont
  {Perra}}, \ and\ \bibinfo {author} {\bibfnamefont {A.}~\bibnamefont
  {Vespignani}},\ }\href {\doibase 10.1371/journal.pone.0022656} {\bibfield
  {journal} {\bibinfo  {journal} {PLoS ONE}\ }\textbf {\bibinfo {volume} {6}},\
  \bibinfo {pages} {e22656} (\bibinfo {year} {2011})}\BibitemShut {NoStop}%
\end{thebibliography}%

\end{document}